\documentclass[twoside,11pt]{idea_style}
\newif\ifdraft
\drafttrue
\usepackage[T1, T2A]{fontenc} % T2A For one russian sentence :(
\usepackage{hyperref}       % hyperlinks
\usepackage{url}            % simple URL typesetting
\usepackage{nicefrac}       % compact symbols for 1/2, etc.
\usepackage{multicol}
\usepackage{microtype}      % microtypography
\usepackage{geometry}
\usepackage{graphicx}%
\graphicspath{{figures/}}
\usepackage{multirow}%
\usepackage{amsmath}
\usepackage{amssymb}
\usepackage{amsfonts}
\usepackage{stmaryrd}
\usepackage{mathrsfs}%
\usepackage{xcolor}%
\usepackage{textcomp}%
\usepackage{manyfoot}%
\usepackage{booktabs}%
\usepackage{algorithm}%
\usepackage{tabularx}%
\usepackage{algorithmicx}%
\usepackage{algpseudocode}%
\usepackage{listings}%
\usepackage{algpseudocode}
\usepackage[inline]{enumitem}
\usepackage{xspace}
\usepackage{tikz}
\usepackage{makecell}
\usepackage{longtable}
\usepackage{changepage}
\usepackage{color}
\usepackage{colortbl}
\usepackage{multirow}
\usepackage{pifont}
\usepackage{subcaption}
\usepackage{physics}
\usepackage[russian,english]{babel}
\usepackage{booktabs}
\usepackage[numbers]{natbib}
\usepackage{setspace}
\usepackage{changepage} % to adjust margins
\usepackage{float}

\usepackage{tablefootnote}
\usepackage{threeparttable}
\usepackage{placeins}

\author[]{Zhiyuan Zhao$^\dagger$, \ Lijian Lin, \ Ye Zhu, \ Kai Xie, \  Yunfei Liu, \  Yu Li}

\affiliation[]{International Digital Economy Academy (IDEA)}
\contribution[\dagger]{Project Lead}

\correspondence{Zhiyuan Zhao \email{zhaozhiyuan@idea.edu.cn}}

\ideadata[Demo page]{\href{https://lemas-project.github.io/LEMAS-Project}{\texttt{https://lemas-project.github.io/LEMAS-Project}}}
\ideadata[Dataset]{\url{https://huggingface.co/LEMAS-Project}}
\ideadata[Code]{\url{https://github.com/LEMAS-Project}}

\title{{LEMAS:}\\[7pt]\Large A 150K-Hour Large-scale Extensible Multilingual Audio Suite with Generative Speech Models}

\abstract{
\small{

Generative speech models have achieved strong performance in high-resource languages such as English and Chinese, yet their capabilities remain limited in multilingual settings. This gap is largely driven by data challenges: collecting and curating large-scale multilingual speech with consistent quality and fine-grained temporal annotations is substantially more difficult, while existing web-crawled corpora provide limited guarantees on data quality and annotation reliability in multilingual settings. To address this bottleneck, we present the \textbf{LEMAS-Dataset}, which, to our knowledge, is currently the largest open-source multilingual speech corpus with word-level timestamps. Covering over 150,000 hours across 10 major languages, LEMAS-Dataset is constructed via a efficient data processing pipeline that ensures high-quality data and annotations. To validate the effectiveness of LEMAS-Dataset across diverse generative paradigms, we train two benchmark models with distinct architectures and task specializations on this dataset. \textbf{LEMAS-TTS}, built upon a non-autoregressive flow-matching framework, leverages the dataset's massive scale and linguistic diversity to achieve robust zero-shot multilingual synthesis. Our proposed accent-adversarial training and CTC loss mitigate cross-lingual accent issues, enhancing synthesis stability. Complementarily, \textbf{LEMAS-Edit} employs an autoregressive decoder-only architecture that formulates speech editing as a masked token infilling task. By exploiting precise word-level alignments to construct training masks and adopting adaptive decoding strategies, it achieves seamless, smooth-boundary speech editing with natural transitions. Experimental results demonstrate that models trained on LEMAS-Dataset deliver high-quality synthesis and editing performance, confirming the dataset's quality. We envision that this richly timestamp-annotated, fine-grained multilingual corpus will drive future advances in prompt-based speech generation systems.
}
}

\begin{document}

\maketitle

\section{Introduction}

\begin{figure}[t]
\centering
    \includegraphics[width=\linewidth]{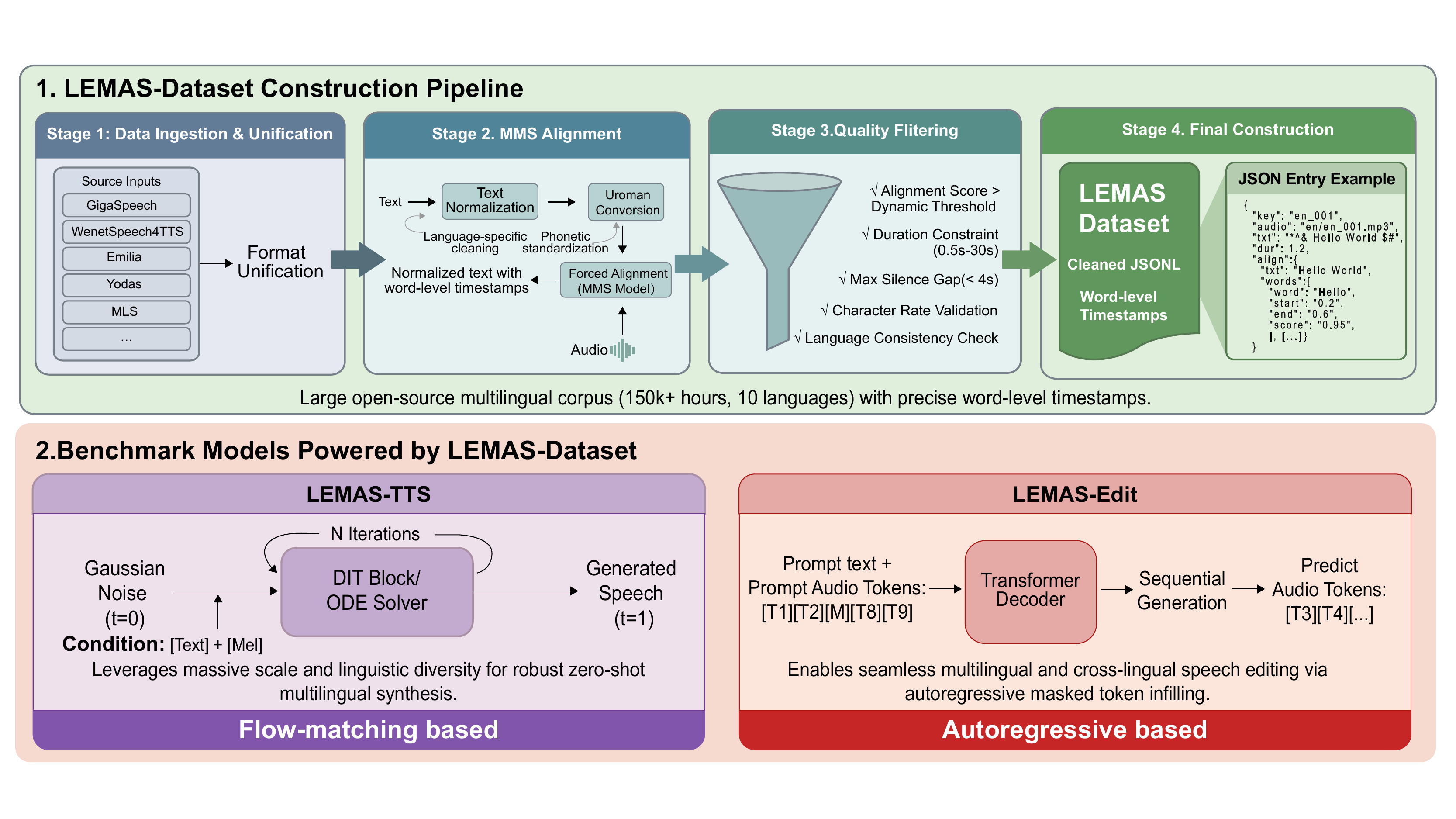}
    \caption{LEMAS-Dataset contains more than $\mathbf{150k}$ hours of multi-speaker speech with forced word-level alignments across \textbf{10} major languages. Based on LEMAS-Dataset, we train two models. \textbf{LEMAS-TTS} performs large-scale, flow-based neural TTS that streams high-fidelity speech from text and a short reference clip, while \textbf{LEMAS-Edit} performs codec-based, word-level speech editing.
}
    \label{fig:teaser}
\end{figure}

The field of speech synthesis has entered a transformative phase with the rise of generative foundation models. Leveraging large-scale pre-training on massive unlabeled audio, recent systems have achieved impressive zero-shot performance, particularly in high-resource languages such as English and Chinese~\cite{du2024cosyvoice, guo2024fireredtts, Wang2024MaskGCTZT, chen2024f5, wang2025spark, indextts2}.
Despite this progress, extending generative speech models to truly multilingual scenarios remains challenging. Compared to English and Chinese, both open-source multilingual datasets and foundation models are still limited in scale, coverage, and annotation quality, leading to noticeable performance degradation across languages. As model capacity continues to grow, the primary bottleneck has increasingly shifted from architecture design to data—specifically, the scarcity of large-scale, high-quality, and fine-grained multilingual speech corpora.

\textbf{Limitations of Existing Speech Datasets.}
While recent efforts have substantially scaled up publicly available speech resources, several fundamental limitations remain. Large-scale corpora such as Libri-Light~\cite{kahn2020libri}, GigaSpeech~\cite{chen2021gigaspeech}, and WenetSpeech~\cite{zhang2022wenetspeech} provide abundant audio–text pairs, enabling the training of robust ASR and TTS models. However, these datasets are predominantly monolingual, focusing on either English or Chinese, and do not address the scarcity of large-scale multilingual resources. Conversely, multilingual datasets such as TEDx~\cite{tedx} and MLS~\cite{pratap2020mls} are constrained by limited scale and narrow domain coverage. Specifically, TEDx consists mainly of public speeches, whereas MLS is restricted to the audiobook domain. More recently, large-scale multilingual collections such as YODAS~\cite{li2023yodas} have been introduced, but they rely heavily on minimally filtered web-crawled data, resulting in highly variable quality and a lack of reliable fine-grained annotations.

We also acknowledge promising efforts such as WenetSpeech4TTS~\cite{ma2024wenetspeech4tts}, which reprocesses WenetSpeech with rigorous cleaning and introduces timestamp-level annotations tailored for speech synthesis. While effective, such datasets remain confined to a single language. In contrast, our goal is to extend this level of data quality and fine-grained temporal supervision to a truly multilingual setting, motivating the construction of LEMAS-Dataset. Besides, built upon LEMAS-Dataset, we design two foundation models:  LEMAS-TTS and LEMAS-Edit (Figure~\ref{fig:teaser}).

\textbf{LEMAS-Dataset.}
To bridge this gap, we introduce LEMAS-Dataset. To the best of our knowledge, LEMAS-Dataset is the largest open-source multilingual speech corpus with rigorous word-level timestamps, comprising over \textbf{150,000 hours} of speech across \textbf{10 major languages}, including Chinese, English, Russian, Spanish, Portuguese, German, French, Italian, Indonesian and Vietnamese. Table~\ref{tab:dataset_comparison} highlights LEMAS-Dataset’s advantages in both scale and temporal alignment precision compared to existing corpora. 

Constructing such a dataset from in-the-wild audio poses unique challenges, as traditional HMM-based forced alignment tools (e.g., MFA~\cite{montreal-forced-aligner}) often struggle with noisy, unsegmented speech. We address this challenge through a robust multi-stage processing pipeline that leverages the MMS aligner~\cite{pratap2023scaling} to extract word-level timestamps and, crucially, assigns a confidence score to each aligned word. This design enables flexible, reliability-aware data filtering, allowing researchers to dynamically trade off data scale and alignment precision according to downstream requirements, and ensuring high utility for precision-critical applications. 

\begin{table}[h]
\caption{Comparison of LEMAS-Dataset with existing large-scale processed speech datasets. Audiobook data refers to professionally recorded, script-aligned speech
with clean acoustic conditions, while in-the-wild data consists of naturally occurring speech collected from real-world scenarios, often exhibiting background noise, spontaneous speaking styles, and imperfect text-audio alignment. LEMAS-Dataset provides extensive total duration and broad language coverage among datasets that include utterance-level timestamps across diverse data sources.}
\label{tab:dataset_comparison}
\centering
\small
\setlength{\tabcolsep}{5pt}
\renewcommand{\arraystretch}{0.95}

\begin{tabular}{ccccc}
\hline
\hline
\textbf{Dataset} & \textbf{Total Duration (h)} & \textbf{Languages} & \textbf{timestamps} & \textbf{Data Source}  \\
\hline
Libri-light~\cite{kahn2020libri}  & 60k & 1  & $\times$  & Audiobook \\
GigaSpeech~\cite{chen2021gigaspeech} & 10k & 1  & $\times$    & Audiobook/In-the-wild\\
GigaSpeech2~\cite{yang2025gigaspeech} & 28k & 3 & $\times$ & Audiobook/In-the-wild \\
WenetSpeech~\cite{zhang2022wenetspeech} & 22k & 1  & $\times$   & In-the-wild\\
WenetSpeech4TTS~\cite{ma2024wenetspeech4tts} & 13k & 1  & \ding{51}  & In-the-wild\\
MLS~\cite{pratap2020mls} & 51k & 8  & $\times$   & Audiobook \\
Emilia~\cite{he2024emilia} & 102k & 6  & $\times$  & In-the-wild \\
Voxbox~\cite{wang2025spark}  & 103k & 2  & $\times$  & Audiobook/In-the-wild \\
\hline
\textbf{LEMAS-Dataset (ours)}  & \textbf{150k} & \textbf{10} & \ding{51}  & \textbf{Audiobook/In-the-wild} \\
\hline
\hline
\end{tabular}
\end{table}

To demonstrate the versatility of LEMAS-Dataset across diverse generative paradigms, we develop two parallel
foundation models, utilizing the dataset’s scale and precision, respectively:

\textbf{LEMAS-TTS.}
Without language-specific alignment regularization, flow-matching models suffer from alignment drift that degrades intelligibility in low-resource languages, as well as accent leakage that replaces target prosody with that of higher-resource languages~\cite{chen2024f5}, ultimately undermining zero-shot cross-lingual fidelity. To address these challenges, we extend the F5-TTS~\cite{chen2024f5} architecture with language-aware stabilization mechanisms. Specifically, we unify diverse writing systems into a shared phonetic space and introduce two complementary objectives: a Connectionist Temporal Classification (CTC) loss~\cite{graves2006ctc} to improve speech intelligibility, and an accent-adversarial loss~\cite{grl} to suppress cross-lingual accent interference. These enhancements enable the model to effectively leverage the full 150k-hour corpus, achieving both high speaker similarity and naturalness in zero-shot cross-lingual synthesis.

\textbf{LEMAS-Edit.}
To further validate the effectiveness of LEMAS-Dataset for autoregressive speech generation and editing, we develop LEMAS-Edit, a specialized editing model built upon the decoder-only architecture. VoiceCraft~\cite{voicecraft} formulates speech editing as a masked token-infilling task, whose generation quality is highly sensitive to boundary precision and decoding stability. Misaligned edit boundaries can lead to noticeable artifacts, such as prolonged silences or unnatural acoustic discontinuities~\cite{voicecraft}. To mitigate repetition artifacts commonly observed during autoregressive decoding, we introduce a repetition penalty at inference time to suppress token looping. We further extend the editing pipeline with system-level components, including denoising, re-alignment using MMS, and consistency verification, enabling precise and robust user-controlled editing. As demonstrated in our demos, LEMAS-Edit achieves seamless, smooth-boundary editing and remains effective on in-the-wild audio with environmental noise or background sounds, highlighting the strong generalization enabled by LEMAS-Dataset.

In summary, our contributions are threefold:

\begin{itemize}
    \item \textbf{LEMAS-Dataset:} We release the largest open-source multilingual speech corpus with rigorous word-level timestamps, comprising over 150,000 hours across 10 languages. The dataset is constructed via a scalable alignment pipeline with word-level confidence estimation, enabling reliable, fine-grained temporal supervision for precision-critical speech generation tasks.
    
    \item \textbf{LEMAS-TTS:} We develop a non-autoregressive flow-matching TTS model for large-scale multilingual training. By unifying scripts into a shared phonetic space and introducing CTC-based alignment regularization and accent-adversarial objectives, LEMAS-TTS achieves robust zero-shot cross-lingual synthesis with improved intelligibility and accent consistency.
        
    \item \textbf{LEMAS-Edit:} We develop an autoregressive decoder only speech editing system with stabilized inference through repetition penalty and speech-rate-based anomaly detection, supported by a robust pipeline with denoising, re-alignment, and verification. LEMAS-Edit enables seamless, smooth-boundary multilingual editing and generalizes well to in-the-wild noisy audios.

\end{itemize}

% These will be the appendices in the summary
\section{Related Work}

\subsection{Large-Scale Speech Corpora}

The evolution of speech datasets has been characterized by a transition from curated, read speech to massive, multi-domain collections harvested from the web. This progression addresses the growing hunger of end-to-end models for data scale and diversity.

\paragraph{Curated and Crowdsourced Foundations.}
Early multilingual corpora like {GlobalPhone}~\cite{schultz2013globalphone} and {IARPA BABEL}~\cite{gales2014speech} set the standard for high-quality, curated recordings, though they were limited in scale. {Mozilla Common Voice}~\cite{ardila2019common} later democratized data collection via crowdsourcing, prioritizing speaker diversity. In the domain of audiobooks, {LibriSpeech}~\cite{panayotov2015librispeech} and {Multilingual LibriSpeech (MLS)}~\cite{pratap2020mls} became the de facto benchmarks for ASR, offering clean read speech suitable for reproducible research.

\noindent\textbf{Massive-Scale Multi-Domain Corpora.}
To overcome the domain limitations of audiobooks, researchers turned to large-scale internet sources. {GigaSpeech}~\cite{chen2021gigaspeech} introduced an evolving, multi-domain English corpus derived from audiobooks, YouTube and podcasts, providing 10,000 hours of high-quality labeled audio. This shift marked a move towards modeling diverse, "in-the-wild" recording conditions. 
In the landscape of Chinese speech, {WenetSpeech}~\cite{zhang2022wenetspeech} (and its recent derivative {WenetSpeech4TTS}~\cite{he2024emilia}) aggregated tens of thousands of hours of multi-domain Mandarin speech, further validating the efficacy of web-mined data for training industrial-grade ASR and TTS models.

\noindent\textbf{Multilingual and Large-Scale unsupervised Speech Corpora.}
Recent efforts have greatly expanded multilingual speech datasets in scale and language coverage, focusing on weakly supervised or unlabeled data.
{VoxPopuli}~\cite{wang2021voxpopuli} offers 400k hours of unlabeled speech in 23 languages from European Parliament recordings.
{YODAS}~\cite{li2023yodas} collects 500k+ hours from 100+ languages on YouTube, using automatic captions with variable quality.
These datasets trade label quality for massive scale, fueling self- and semi-supervised speech research.

\subsection{Generative Speech Models}

Neural TTS has shifted from task-specific models to unified generative frameworks. It evolved from autoregressive models like Tacotron~\cite{tacotron} to non-autoregressive FastSpeech~\cite{fastspeech}, and then to end-to-end variational models like VITS~\cite{vits}. Today’s frontier features large-scale foundation models unifying synthesis, cross-lingual cloning, and editing.

\noindent\textbf{Generative Foundation Models.}
Speech generation falls into discrete, continuous, and hybrid paradigms, differing in representation and generation methods.
\begin{itemize}
\item \textbf{Discrete approaches} model speech as sequences of codec tokens autoregressively or via masked infilling, similar to language models. VALL-E~\cite{valle} pioneered zero-shot TTS using EnCodec tokens; AudioLM~\cite{audiolm} adds hierarchical semantic and acoustic token prediction. XTTS~\cite{Casanova2024XTTSAM} extends this to many languages. Recent works focus on efficiency and control, e.g., MaskGCT~\cite{Wang2024MaskGCTZT} and FishAudio~\cite{fishaudio}.
\item \textbf{Continuous approaches} generate mel-spectrograms or waveforms directly from noise using diffusion or flow matching. NaturalSpeech 2~\cite{shen2023naturalspeech2} demonstrates high-fidelity latent diffusion; VoiceBox~\cite{le2023voicebox} simplify pipelines by dropping external encoders. F5-TTS~\cite{chen2024f5} employs a DiT structure with fill-in-the-middle, achieving competitive zero-shot synthesis with a streamlined design.
\item \textbf{Hybrid approaches} combine discrete semantic modeling with continuous acoustic refinement for diverse and faithful synthesis. CosyVoice 1–3~\cite{cosyvoice,du2024cosyvoice2,Du2025CosyVoice3T} merge LLM-based semantic tokens with conditional flow matching for multilingual instruction-following TTS; FireRed-TTS~\cite{Xie2025FireRedTTS2TL} cascades discrete and flow-based models for prosody control. IndexTTS 2~\cite{indextts2} further explores semantic compression and style manipulation efficiently.
\end{itemize}

\noindent\textbf{Multilingual Extensions.}
With growing training corpora, TTS shifted from mono/bilingual to multilingual. Early open-source models like XTTS-v2~\cite{Casanova2024XTTSAM} trained on 17 languages jointly. Community models such as OpenAudio-S1-mini~\cite{fishaudio} support 13 languages with 100k+ hours, enhancing cross-lingual generalization. At a larger scale, CosyVoice~3~\cite{Du2025CosyVoice3T} uses a speech tokenizer from a large audio understanding model, training on over a million hours across 9 languages and 18 Chinese dialects. This reflects rapid data scaling and demand for massive multilingual datasets.

\noindent\textbf{Text-Based Speech Editing.}
Speech editing modifies spoken content while preserving timbre and is now seen as a TTS subtask. VoiceBox~\cite{le2023voicebox} shows versatile in-context learning for zero-shot TTS, editing, style conversion, and diverse generation. Non-autoregressive MaskGCT~\cite{Wang2024MaskGCTZT} uses mask-and-predict learning; F5-TTS~\cite{chen2024f5} employs a Diffusion Transformer (DiT) supporting TTS and editing jointly. Autoregressive VoiceCraft~\cite{voicecraft} combines causal masking and delayed stacking for flexible generation and diverse editing within sequences.
\section{LEMAS-Dataset}

We introduce LEMAS-Dataset, a massive open-source multilingual speech corpus with over 150,000 hours spanning 10 major languages: Chinese (Zh), English (En), Russian (Ru), Spanish (Es), Indonesian (Id), German (De), Portuguese (Pt), Vietnamese (Vi), French (Fr), and Italian (It). LEMAS-Dataset provides word-level alignments along with confidence scores for each word. This fine-grained annotation facilitates a wide range of applications, including data quality filtering based on confidence scores, training duration prediction models, and generating detailed prompts to guide speech synthesis and understanding.

\subsection{Dataset Processing Pipeline}

Constructing a large-scale multilingual speech corpus for generative modeling requires more than merging existing datasets, which vary widely in annotations, transcription quality, and acoustics. The key insight of the LEMAS-Dataset pipeline is to separate \emph{language diversity} from \emph{annotation fragility}, retaining only samples with reliably grounded temporal structure. To achieve this, we develop a multi-stage pipeline that unifies heterogeneous sources, performs robust multilingual alignment, and applies scalable quality control.

\noindent\textbf{Data Ingestion and Structural Unification.}
We first aggregate audio–text pairs from a diverse collection of publicly available speech corpora, including GigaSpeech (English)~\cite{chen2021gigaspeech}, GigaSpeech2 (Indonesian, Vietnamese)~\cite{yang2025gigaspeech}, WenetSpeech4TTS (Chinese, VoiceBox-based)~\cite{ma2024wenetspeech4tts, wang2025spark}, Emilia (Chinese and English, VoiceBox-based)~\cite{he2024emilia}, MLS (English, German, French, Portuguese, Spanish, Italian)~\cite{pratap2020mls}, multilingual TEDx (German, French, Portuguese, Spanish, Italian)~\cite{tedx}, Alcaim (Portuguese)~\cite{alcaim}, Golos (Russian)~\cite{golos}, and Yodas (German, French, Spanish, Portuguese, Russian, Italian, Indonesian, Vietnamese)~\cite{li2023yodas}. These sources span a wide range of recording scenarios, including audiobooks, podcasts, and in-the-wild conversational speech. Rather than preserving dataset-specific schemas, we normalize all inputs into a unified data representation, enabling language-agnostic downstream processing and large-scale automation.

\noindent\textbf{Multilingual MMS Alignment via Romanized Text.}
Reliable word-level timestamps are a prerequisite for both TTS training and speech editing tasks, yet traditional forced alignment pipelines often rely on language-specific lexicons and G2P systems, which are brittle under multilingual, noisy, and code-switched conditions. To avoid these limitations, we adopt the Multilingual MMS Forced Aligner~\footnote{https://docs.pytorch.org/audio/2.3/tutorials/forced\_alignment\_for\_multilingual\_data\_tutorial.html}, a wav2vec-based CTC alignment model~\cite{baevski2020wav2vec, graves2006ctc} provided by torchaudio~\footnote{https://github.com/pytorch/audio}. This model was trained on over 23,000 hours of speech spanning more than 1,100 languages as part of the ``Scaling Speech Technology to 1,000+ Languages'' initiative~\cite{pratap2023scaling}. 

Prior to alignment, transcripts are language-specifically normalized and then romanized using tools such as Uroman~\footnote{https://github.com/isi-nlp/uroman}, which maps diverse scripts (e.g., Chinese characters, Cyrillic) into a shared Latin representation. This design removes the dependency on language-specific pronunciation dictionaries and allows the alignment model to robustly handle rare words, named entities, and intra-sentential code-switching. The MMS aligner then produces word-level boundaries along with token-level posterior probabilities, which form the basis for subsequent quality control.

\noindent\textbf{Confidence-Driven Quality Filtering.}
A central principle of our pipeline is that \emph{successful alignment alone is insufficient}; only alignments with high temporal reliability are suitable for generative speech modeling. We therefore retain only samples for which word-level alignments can be extracted, and apply a series of rule-based filters grounded in alignment confidence and temporal consistency:
\begin{itemize}
    \item \textbf{Alignment Confidence Filtering.}
    For each utterance, we compute an average alignment confidence score based on the posterior probabilities of aligned tokens. To accommodate variability across languages and source corpora, we apply dataset-specific thresholds $x \in [0.2, 0.5]$, retaining only samples whose confidence exceeds the corresponding threshold.

    \item \textbf{Duration and Pause Constraints.}
    Utterances shorter than 0.5 seconds or longer than 30 seconds are removed to ensure compatibility with TTS training. Additionally, samples containing prolonged unaligned regions or silence intervals exceeding 4 seconds are discarded, as such patterns often indicate segmentation or transcription errors.

    \item \textbf{Speech Rate Normalization.}
    To exclude abnormally fast or slow speech, we enforce a language-specific constraint on the ratio between textual length and audio duration. For each language $L$, the character-based ratio $\text{len(text)} / \text{duration}$ must fall within a predefined interval $[\text{min\_ratio}_L, \text{max\_ratio}_L]$, empirically determined from corpus statistics.

    \item \textbf{Language and Character Validation.}
    We restrict the dataset to ten target languages: Chinese (zh), English (en), Russian (ru), Spanish (es), Indonesian (id), German (de), Portuguese (pt), Vietnamese (vi), French (fr), and Italian (it). All transcripts are verified to belong to one of these languages, and samples containing unsupported scripts, emojis, or excessive non-linguistic symbols are removed according to language-specific character sets.
\end{itemize}

All thresholds are selected empirically to balance data scale and alignment precision. Notably, we do not apply explicit speech enhancement or background noise removal during preprocessing. While many TTS systems remain sensitive to acoustic interference, we intentionally preserve in-the-wild characteristics, with the expectation that more advanced generative models will be better equipped to leverage such variability. Consistent with this design choice, examples on our demo page show that the proposed editing model remains effective on noisy recordings and can, to some extent, reproduce acoustic environments consistent with the reference audio, highlighting the practical benefits of retaining realistic background conditions.

\noindent\textbf{Final Representation and Storage.}
The resulting dataset is stored in a unified JSON-based format, containing a unique sample key, audio file paths, original and normalized transcripts, and word-level timestamps with associated confidence scores. For efficient storage and distribution, all audio files are converted to MP3 format while preserving their original sampling rates.

\begin{figure}[h]
\centering
    \includegraphics[width=0.95\linewidth]{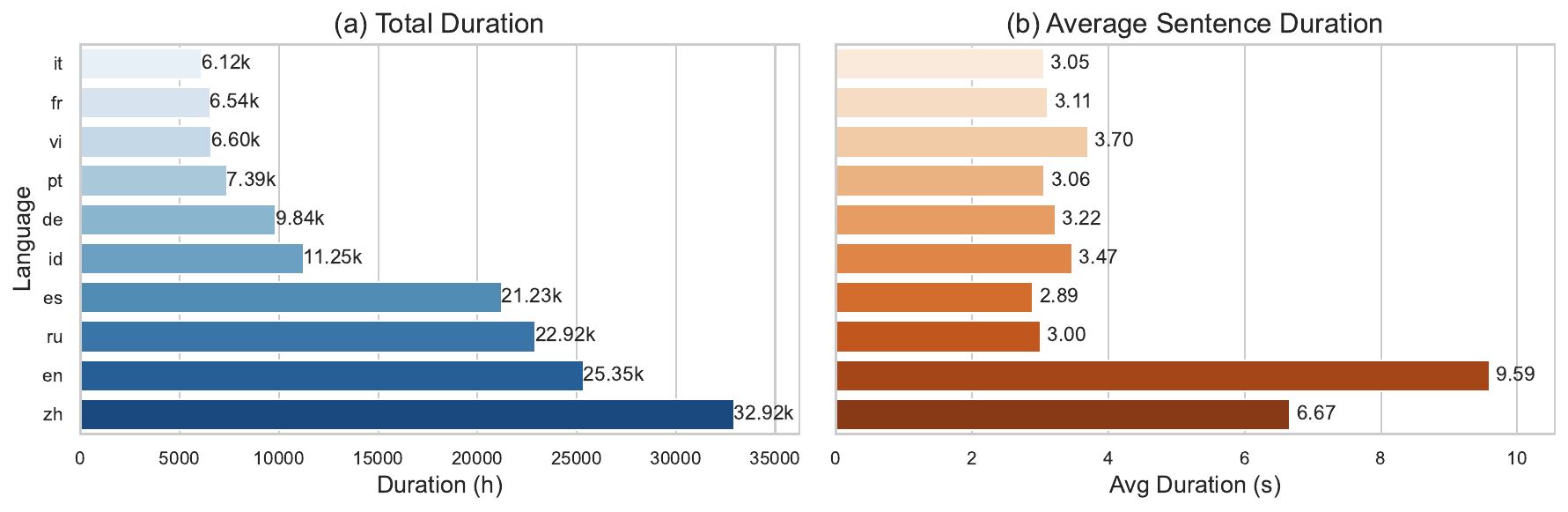}
    \caption{Dataset Statistics of LEMAS-Dataset. (a) Language-wise duration, (b) the average sentence duration in seconds. The dataset shows substantial variation in both data volume and sentence length across languages.}
    \label{fig:dataset_statistics}
\end{figure}

\subsection{Dataset Statistics}

The LEMAS-Dataset consists of approximately 150,144 hours of audio data spanning 10 major languages. As illustrated in Figure~\ref{fig:dataset_statistics}, we explicitly curate the dataset to balance speech duration across languages, resulting in a relatively uniform distribution despite differences in source data availability. While high-resource languages such as Chinese (Zh) and English (En) remain slightly larger, all languages are represented at a substantial scale. In particular, even the lower-resource subsets (e.g., Italian and Vietnamese) exceed 6,000 hours, which enables effective monolingual training as well as joint multilingual modeling. To facilitate reproducible research and consistent benchmarking, the dataset is partitioned into training (train) and evaluation (eval) subsets.

\noindent\textbf{Training Set.} 
Detailed statistics for the training partition are provided in Table~\ref{tab:training_set}. The training set contains over 150,000 hours of audio in total, designed to support the training of large-scale acoustic models.
We note that sentence-level statistics vary across languages due to differences in source corpora. Specifically, Chinese and English data are primarily derived from GigaSpeech, Emilia, and WenetSpeech4TTS, which contain longer-form utterances,
whereas data for other languages are mainly sourced from YODAS, which consists of shorter, pre-segmented utterances without additional merging or restructuring.
As a result, the average utterance length in Chinese and English is noticeably longer than in other languages.

\begin{table}[h]
\caption{Detailed statistics of the \textbf{training} set from LEMAS-Dataset. The units \textbf{h}, \textbf{s}, and \textbf{M} denote hours, seconds, and millions ($10^6$), respectively. Avg. Duration and Avg. Words indicate average duration and average words per utterance.}
\label{tab:training_set}
\centering
\small
\setlength{\tabcolsep}{5pt}
\renewcommand{\arraystretch}{0.95}

\begin{tabular}{cccccc}
\hline
\hline
\textbf{Language} & \textbf{Total Duration (h)} & \textbf{Avg. Duration (s)} & \textbf{Utterances (M)} & \textbf{Total Words (M)} & \textbf{Avg. Words} \\
\hline
it & 6.12k  & 3.05 & 7.21  & 48.86  & 6.78 \\
fr & 6.54k  & 3.11 & 7.56  & 65.84  & 8.71 \\
vi & 6.60k  & 3.70 & 6.43  & 89.90  & 13.99 \\
pt & 7.38k  & 3.06 & 8.68  & 60.32  & 6.95 \\
de & 9.84k  & 3.22 & 11.00 & 80.12  & 7.28 \\
id & 11.25k & 3.47 & 11.66 & 85.97  & 7.37 \\
es & 21.22k & 2.89 & 26.41 & 183.67 & 6.96 \\
ru & 22.92k & 3.00 & 27.47 & 163.02 & 5.93 \\
en & 25.35k & 9.59 & 9.52  & 268.68 & 28.24 \\
zh & 32.92k & 6.67 & 17.78 & 496.96 & 27.96 \\
\hline
total & 150.14k & 4.18 & 133.71  & 1,543.34    & 12.02    \\
\hline
\hline
\end{tabular}
\end{table}

\begin{table}[h]
\centering
\caption{Detailed statistics of \textbf{evaluation} set from LEMAS-Dataset. The units \textbf{min} and \textbf{s} denote minutes and seconds, respectively. Avg. Duration and Avg. Words indicate average duration and average words per utterance.}
\label{tab:validation_set}
\small
\setlength{\tabcolsep}{5pt}
\renewcommand{\arraystretch}{0.95}

\begin{tabular}{cccccc}
\hline
\hline
\textbf{Language} & \textbf{Total Duration (min)} & \textbf{Avg. Duration (s)} & \textbf{Utterances} & \textbf{Total Words} & \textbf{Avg. Words} \\
\hline
it & 44.22      & 5.31  & 500   & 6,599    & 13.20  \\
fr & 38.17      & 4.58   & 500   & 6,546    & 13.09 \\
vi & 36.74      & 4.41  & 500   & 6,727    & 13.45 \\
pt & 41.69      & 5.00  & 500   & 5,812    & 11.62 \\
de & 38.65      & 4.64  & 500   & 5,599    & 11.20  \\
id & 47.20       & 5.67  & 500   & 6,133    & 12.27 \\
es & 40.52      & 4.86  & 500   & 6,216    & 12.43 \\
ru & 40.24      & 4.82  & 500   & 5,138    & 10.28 \\
en & 67.46      & 8.10  & 500   & 11,325   & 22.65 \\
zh & 75.84      & 9.10  & 500   & 18,627   & 37.25 \\
\hline
total   & 470.73     & 5.65 & 5,000  & 78,722 & 15.74 \\
\hline
\hline
\end{tabular}
\end{table}

\noindent\textbf{Evaluation Set}. 
To establish a consistent benchmark, we curate a separate evaluation set, as detailed in Table~\ref{tab:validation_set}. Unlike the training set, the evaluation set is balanced by strictly selecting 500 utterances per language based on rigorous quality criteria. Specifically, we filter for samples with an average word-level alignment score greater than 0.9, a word count exceeding 5, and a duration between 3 and 15 seconds. Additionally, sentence-end silence is trimmed to a maximum of 0.2\,s to standardize acoustic characteristics. From the pool of samples that satisfy the above conditions, we further compute language-specific statistics of the character-to-duration ratio and select the 500 utterances closest to the per-language mean. This additional step helps exclude atypical or anomalous cases, such as singing or exaggerated speaking styles, while preserving representative speaking patterns for each language. Finally, to further ensure data quality, approximately 20\% of the evaluation samples were manually reviewed to verify that the test set consists exclusively of natural speech.
This curation ensures that evaluation metrics (e.g., WER or MOS) are comparable across different languages without being biased by data volume differences or quality variations.

\section{LEMAS-TTS}

We design a TTS method, LEMAS-TTS, builds upon the F5-TTS architecture~\cite{chen2024f5}, utilizing its non-autoregressive Flow Matching framework and Diffusion Transformer (DiT) backbone. While F5-TTS excels at zero-shot tasks, scaling it to a robust multilingual model presents challenges such as alignment stability and effective prosody transfer across languages. To overcome these, we introduce architectural improvements that unify linguistic representations and stabilize the training process.

\subsection{Multilingual Training Framework}

To facilitate high-fidelity multilingual synthesis, we revise the training paradigm to include a unified phonetic input space, auxiliary supervision objectives, and fine-grained prosody modeling.

\noindent\textbf{Unified Phonetic Representation.} To bridge the linguistic gap across diverse languages, we depart from the mixed character/word-level tokenization of the original system. Instead, we adopt a unified phonetic front-end. For Chinese, we employ an initial-final Pinyin representation with tones to capture syllabic structure; for other languages, we utilize the International Phonetic Alphabet (IPA) sequences derived via eSpeak-NG \cite{espeakng}. We choose IPA for these languages because eSpeak-NG’s frontend can automatically handle language switching, and using IPA also facilitates verification of correct pronunciation. All phonetic tokens are augmented with explicit language identifiers (e.g., \texttt{<zh>}, \texttt{<en>}) to project distinct languages into a shared latent space while preserving language-specific phonotactics. Additionally, we leverage word-level timestamps to insert explicit pause tags (\texttt{\#1}–\texttt{\#4}) within the phoneme sequences to denote short, medium, long and abnormal pauses, enabling simple yet effective pause control.

\noindent\textbf{Auxiliary Stabilization Objectives.}
Standard flow-matching objectives can sometimes face convergence difficulties when dealing with complex multilingual scenarios. To address this, we introduce two auxiliary loss functions to improve alignment accuracy and speech intelligibility:
\begin{itemize}
    \item \textbf{CTC Alignment Loss.} To enforce strict monotonicity between the acoustic trajectory and linguistic content, we apply a Connectionist Temporal Classification (CTC) loss on the decoder outputs. This is implemented via a lightweight projection head ~\cite{graves2006ctc} that maps the predicted mel-spectrograms to phone sequences, explicitly encouraging accurate duration modeling.
    \item \textbf{Accent-Adversarial Disentanglement.} Inspired by IndexTTS2~\cite{indextts2}, we aim to disentangle speaker timbre from accent information. We attach an accent classifier to the conditioning pathway and optimize it via a Gradient Reversal Layer (GRL). By supervising this classifier with pseudo-labels from an off-the-shelf language identification model, the encoder is forced to learn accent-invariant representations, thereby reducing cross-lingual accent leakage.
\end{itemize}

\noindent\textbf{Cross-Lingual Prosody Modeling.}
To capture fine-grained prosodic variations, we integrate a Prosody Encoder adapted from the Seamless Expressive architecture~\cite{SeamlessM4TArXiv}. Specifically, we extract 80-bin filterbank features from segmented sub-utterances and encode them into a 512-dimensional prosody embedding using pretrained ECAPA-TDNN backbone. These embeddings are aligned with both mel-frames and text tokens and injected into the DiT via linear projection layers. This mechanism provides explicit prosodic supervision, significantly improving the naturalness of prosody transfer across languages.

\noindent\textbf{Design Analysis and Exclusions.}
During the development of LEMAS-TTS, we also explored explicit speaker contrastive learning (InfoNCE) and clip-and-shuffle data augmentation. However, empirical observations indicated that enforcing strict speaker invariance via InfoNCE conflicted with the flow-matching objective, leading to reduced stability. Similarly, shuffling segmented mel-chunks disrupted the temporal coherence required for the DiT to learn consistent continuations. Consequently, these components were excluded from the final architecture in favor of the alignment and prosody objectives described above.

\subsection{Inference Pipeline and Sampling Dynamics}

To ensure robust multilingual generation, our inference framework strictly aligns with the training paradigm to minimize distribution shift, while incorporating an adjustable sampling strategy that enables flexible control over sampling dynamics to enhance stability across varying acoustic conditions and speech styles.

\noindent\textbf{Unified Multilingual Front-End.}
We deploy a standardized text processing pipeline that mirrors the training stage to ensure input consistency. Firstly, The system leverages automatic language identification to apply language-specific normalization rules (e.g., number expansion and punctuation filtering).
Then, text is converted into the shared phonetic space defined in our training strategy: Chinese is mapped to tonal Pinyin with initial--final decomposition, while other languages are converted into IPA sequences. Crucially, the system supports explicit insertion of spaces and pause tokens (\texttt{\#1-\#3}) within the text, enabling precise control over pause durations.

\begin{figure}[h]
\centering
    \includegraphics[width=0.95\linewidth]{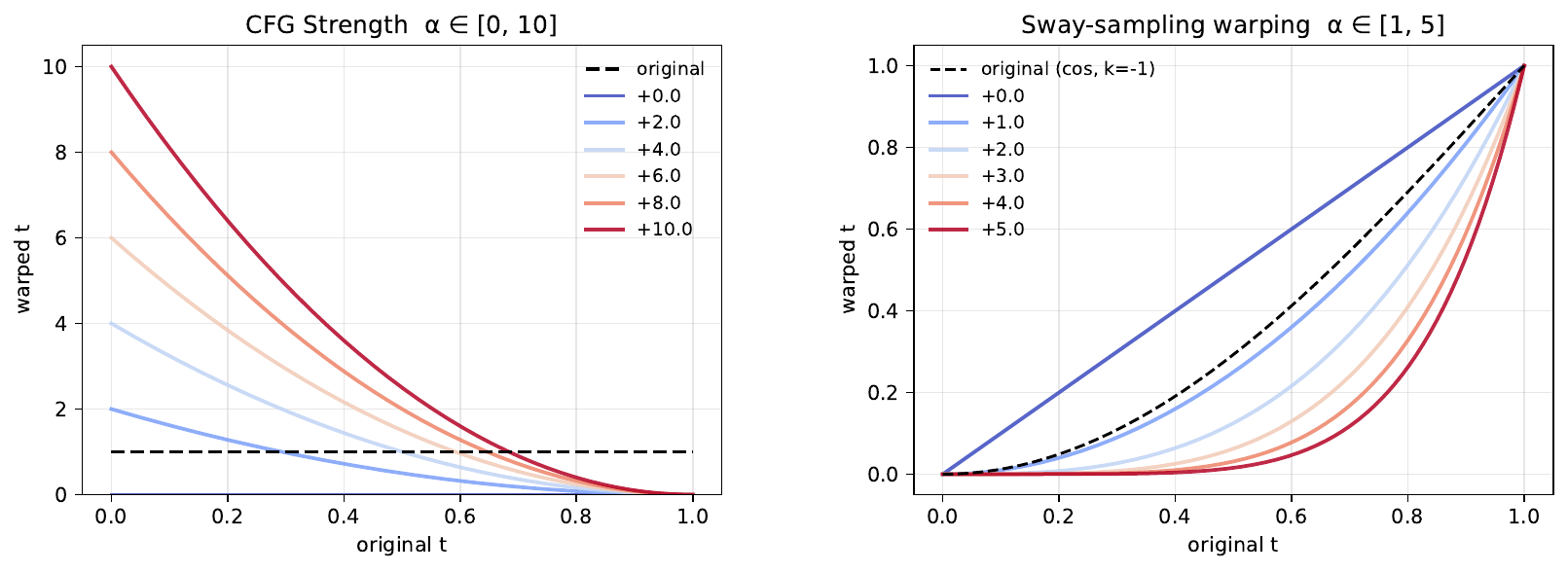}
    \caption{Time-dependent control schedules used in our sampling strategy. 
    Left: CFG strength schedules with different maximum guidance levels, emphasizing early timesteps and gradually decaying over sampling. 
    Right: Sway-sampling time warping with varying strengths, compared to a cosine-based baseline (dashed), where stronger warping allocates more steps to later timesteps. Together, these schedules demonstrate how guidance strength and sampling allocation can be shaped over time to influence generation behavior.}
    \label{fig:sway_cfg}
\end{figure}

\noindent\textbf{Dynamic Guidance and Sway Sampling.}
The original Sway Sampling strategy in F5-TTS allocates more steps in the early generation phase to improve coarse structural accuracy. However, it performs suboptimally in multilingual and cross-lingual scenarios. To address this, we redesign the time-warping function and classifier-free guidance (CFG) schedule as one-parameter families, as illustrated in Figure~\ref{fig:sway_cfg}, enabling smooth interpolation between linear and non-linear behaviors. By increasing early-stage steps and CFG scale, we improve pronunciation accuracy, while reducing CFG strength later allows greater model flexibility for enhanced speaker similarity and audio quality.

nificantly enhancing perceptual stability for long, multilingual utterances.

First, we refine the Classifier-Free Guidance (CFG) mechanism. Let $f_{\theta}^{\mathrm{cond}}(x,t)$ and $f_{\theta}^{\mathrm{uncond}}(x,t)$ denote the conditional and unconditional flows, respectively. We define the guided flow as:
\begin{equation}
f_{\theta}^{\mathrm{cfg}}(x,t)
\;=\;
f_{\theta}^{\mathrm{cond}}(x,t)
\;+\;
g(t)\,\bigl(f_{\theta}^{\mathrm{cond}}(x,t) - f_{\theta}^{\mathrm{uncond}}(x,t)\bigr),
\label{eq:cfg}
\end{equation}
where the time-dependent schedule $g(t)$ is defined to decay quadratically as:
\begin{equation}
g(t) \;=\; \lambda\,(1 - t)^{2},
\label{eq:guidance_schedule}
\end{equation}
with $\lambda$ being a scaling factor (e.g., $\lambda = 5$). This schedule applies stronger structural correction during the initial high-noise regime ($t \approx 0$) and gradually relaxes guidance near $t=1$ to preserve acoustic naturalness, as shown in Equations~\eqref{eq:cfg} and \eqref{eq:guidance_schedule}.

Second, to further optimize the discretization of the flow trajectory, we update the Sway Sampling strategy. Instead of a linear or cosine-based grid, we utilize a power-law reparameterization as Equation~\eqref{eq:powerlaw_reparam} represents. Given a uniform grid $s_k \in [0,1]$ for $k=0,\dots,K$, the effective time steps are defined as:
\begin{equation}
t_k \;=\; s_k^{\,1 + \gamma},
\label{eq:powerlaw_reparam}
\end{equation}
where $\gamma \ge 0$ is a tunable curvature parameter. Increasing $\gamma$ effectively stretches the early sampling steps and compresses the later ones. Empirically, this redistribution of computational effort improves the recovery of signal from noise, significantly enhancing perceptual stability for long, multilingual utterances.

\noindent\textbf{Versatile Conditioning Modes.}  
The inference interface is extended to support flexible control over prosody and style. By enabling the prosody encoder, users can extract a global prosody embedding from a reference waveform. This embedding is injected into both the mel and text conditioning streams to facilitate precise style transfer. Additionally, we introduce a reference-free mode that discards conditioning on reference audio, allowing the model to generate speech with randomly sampled speaker characteristics and more native-like pronunciation. This encourages the model to rely on internal priors for timbre and energy, enabling pure text-driven synthesis.

\section{LEMAS-Edit}

We develop LEMAS-Edit by significantly extending the architecture of VoiceCraft~\cite{voicecraft}, transforming it from a monolingual English model into a multilingual speech editing framework. Our contributions encompass an expansion of training data, a re-engineered distributed training framework, and the introduction of adaptive decoding strategies to ensure generation stability.

\subsection{Multilingual Adaptation and Training}

Crucially, when the language context switches within a sentence, new language tags are inserted dynamically. This structure allows the autoregressive model to switch its internal linguistic context mid-utterance, naturally supporting mixed-lingual editing tasks.

We have re-engineered the training infrastructure to support large-scale distributed learning. The framework now handles data sharding across multiple heterogeneous datasets and GPU ranks, ensuring balanced data consumption. A significant optimization is the {simplification of the optimization step}. In the original implementation, gradient accumulation was used to simulate larger batches, which added complexity to the training dynamics. We remove this mechanism entirely in favor of single-step updates on full batches. This modification not only simplifies the codebase but also makes the relationship between batch size and effective gradient updates transparent, enhancing training stability and robustness against memory fragmentation when processing long multilingual sequences on large GPU clusters.

\subsection{Inference and Decoding Strategy}

\noindent\textbf{History-Aware Repetition Control.}
Autoregressive codec models often suffer from repetition loops, especially in long-form generation. To mitigate this, we introduce a dynamic, history-aware penalty mechanism within the \texttt{top-k/top-p} sampling process. Unlike static penalties, our approach scales with the generation length. We modify the logits of previously generated tokens based on their frequency using the following formulation:
\begin{equation}
  \text{penalty} = \frac{\text{repetition\_penalty}}{100} \cdot \text{num\_gen} + 1.
  \label{eq:rep_penalty}
\end{equation}
Here, \texttt{num\_gen} represents the current number of generated tokens. For any token present in the history, its positive logits are divided by this penalty, while negative logits are multiplied by it. This progressively increases the cost of repeating tokens as the sequence grows, effectively suppressing both silence loops and content stuttering without degrading the initial prompt adherence.

\noindent\textbf{Adaptive Re-Generation Mechanism.}
To ensure that the generated audio duration aligns with natural speech patterns, we implement a multi-round adaptive inference scheme. First, we establish a baseline speaking rate by analyzing the reference audio:
\begin{equation}
  \texttt{avg\_speed} = \frac{\#\text{encodec frames}}{\#\text{raw text tokens}}.
  \label{eq:avg_speed}
\end{equation}
This estimated rate allows us to impose a dynamic \texttt{max\_rate} constraint on the decoder, preventing run-away generation.

However, simple constraints are insufficient for complex edits. Therefore, we introduce a {feedback loop}. During generation, if the model triggers an internal anomaly flag (\texttt{RE\_GEN}) or if the output length is pathologically short (specifically, $< 0.5 \times \texttt{avg\_speed}$), the system automatically triggers a re-generation pass. In each subsequent round, we iteratively relax the constraints by expanding the editing mask boundaries (providing more acoustic context) and incrementing the \texttt{repetition\_penalty}. This iterative process continues until the output matches the expected prosodic structure or a maximum retry limit is reached.

\subsection{Multilingual Editing Front-End}

We have upgraded the inference front-end to support a professional-grade multilingual workflow, focusing on accuracy and flexibility:

\noindent\textbf{Robust Recognition and Alignment.} Precise text-to-audio alignment is essential for accurate speech editing. To enhance transcription robustness, we replace the standard English ASR with the multilingual Whisper \texttt{large} model. Additionally, we employ the MMS Forced Aligner to obtain more accurate word-level timestamps. This fine-grained temporal information is crucial for constructing precise editing masks, ensuring modifications are strictly confined to the intended regions without affecting adjacent words.

\noindent\textbf{Signal Enhancement.} To handle real-world recordings, the pipeline integrates two distinct denoising backends. Users can select UVR5\footnote{https://ultimatevocalremover.com} for mild denoising that preserves background ambience (crucial for maintaining naturalness in edited speech), or DeepFilterNet\footnote{https://github.com/Rikorose/DeepFilterNet} for aggressive suppression of noise in low-quality inputs.

\noindent\textbf{Long-Form Processing.} Processing extended recordings in a single pass often degrades attention performance. Our system automatically segments long audio files into manageable chunks, processes the edits independently, and stitches the results back together using smooth cross-fades at zero-crossing points. This ensures perceptual continuity across the entire recording.

\noindent\textbf{Unified Interface.} The system exposes two distinct operational modes within a single interface: \textit{TTS Mode}, which generates speech autoregressively starting from the end of the sentence backwards using reference text and audio, and \textit{Editing Mode}, which modifies existing audio. The latter leverages the LEMAS-Dataset's precise alignment capabilities to allow granular, human-in-the-loop adjustments of start and end times, bridging the gap between automated models and professional editing requirements.

\section{Experiment}

We construct a unified multilingual benchmark based on the LEMAS-Dataset to evaluate both text-to-speech synthesis and text-based speech editing. The benchmark targets pronunciation stability, cross-lingual generalization, and word-level editing naturalness, leveraging the dataset’s fine-grained word-level timestamps. For multilingual TTS evaluation, we randomly sample 10 utterances per language across 10 languages and synthesize speech using all target texts, yielding 10,000 samples that cover both intra- and cross-lingual settings. Objective evaluation is conducted using Word Error Rate (WER) and speaker similarity. For speech editing, we construct word-level editing cases by randomly replacing consecutive words in aligned utterances across 7 languages and evaluate naturalness via subjective A/B tests, focusing on boundary smoothness and perceptual coherence after localized edits. All evaluations follow a unified data construction and inference protocol to ensure fair comparison across models and tasks. Evaluation lists and generation scripts will be released for reproducibility.

\subsection{Evaluation of LEMAS-TTS}

We evaluate LEMAS-TTS on 10 languages and compare it with the contemporaneous open-source baseline OpenAudio-S1-mini\footnote{https://huggingface.co/spaces/fishaudio/openaudio-s1-mini}, which supports the same language set. Word error rate (WER) is evaluated using FunASR Paraformer-zh\footnote{https://huggingface.co/funasr/paraformer-zh} for Chinese and Whisper-large-v3\footnote{https://huggingface.co/openai/whisper-large-v3} for all other languages, while speaker similarity (SIM) is computed with WavLM-large\footnote{https://huggingface.co/microsoft/wavlm-large}, following the Seed-TTS evaluation protocol~\cite{Anastassiou2024SeedTTSAF}. Due to the lack of native evaluators for languages other than Chinese and English, no subjective evaluation was conducted.

\begin{table}[!t] 
\centering
% \vspace{-2cm}
% \renewcommand{\arraystretch}{1.2}
\caption{Objective evaluation of TTS models on multilingual test sets.  
WER (Word Error Rate) measures transcription accuracy (lower is better), while SIM (Similarity) evaluates perceptual closeness to the target speaker (higher is better).  
Results compare OpenAudio-S1-Mini and LEMAS-TTS models, with and without prosody control, across 10 languages.}\label{tab:cmp_wer}
\begin{tabular}{cccc}
\hline
\hline
\textbf{Language} & \textbf{Model} & \textbf{WER↓(\%)} & \textbf{SIM↑} \\
\hline
\multirow{3}{*}{de} 
 & OpenAudio-S1-Mini                 & 3.95 & 0.448 \\
 & LEMAS-TTS w/o prosody             & 1.20 & 0.533 \\
 & LEMAS-TTS w/ prosody              & \textbf{0.44} & \textbf{0.535} \\
\hline
\multirow{3}{*}{en} 
 & OpenAudio-S1-Mini                 & 4.98 & 0.451 \\
 & LEMAS-TTS w/o prosody             & 2.21 & \textbf{0.557} \\
 & LEMAS-TTS w/ prosody              & \textbf{1.07} & 0.556 \\
\hline
\multirow{3}{*}{es} 
 & OpenAudio-S1-Mini                 & 7.66 & 0.488  \\
 & LEMAS-TTS w/o prosody             & 4.93 & \textbf{0.560}  \\
 & LEMAS-TTS w/ prosody              & \textbf{4.28} & 0.552  \\
\hline
\multirow{3}{*}{fr} 
 & OpenAudio-S1-Mini                 & 13.41 & 0.458  \\
 & LEMAS-TTS w/o prosody             & 10.45 & 0.517  \\
 & LEMAS-TTS w/ prosody              & \textbf{9.87} & \textbf{0.518}   \\
\hline
\multirow{3}{*}{id} 
 & OpenAudio-S1-Mini                 & 32.77 & 0.518  \\
 & LEMAS-TTS w/o prosody             & 7.05 & \textbf{0.550}   \\
 & LEMAS-TTS w/ prosody              & \textbf{6.38} & 0.534   \\
\hline
\multirow{3}{*}{it} 
 & OpenAudio-S1-Mini                 & 9.27 & 0.489  \\
 & LEMAS-TTS w/o prosody             & 5.57 & \textbf{0.544}  \\
 & LEMAS-TTS w/ prosody              & \textbf{4.43} & 0.537  \\
\hline
\multirow{3}{*}{pt} 
 & OpenAudio-S1-Mini                 & 8.05 & 0.476  \\
 & LEMAS-TTS w/o prosody             & 3.12 & \textbf{0.555}  \\
 & LEMAS-TTS w/ prosody              & \textbf{2.09} & 0.546  \\
\hline
\multirow{3}{*}{ru} 
 & OpenAudio-S1-Mini                 & 20.75 & 0.475  \\
 & LEMAS-TTS w/o prosody             & 12.29 & \textbf{0.545}  \\
 & LEMAS-TTS w/ prosody              & \textbf{11.84} & 0.538  \\
\hline
\multirow{3}{*}{vi} 
 & OpenAudio-S1-Mini                 & - & 0.451 \\
 & LEMAS-TTS w/o prosody             & 22.14 & \textbf{0.541} \\
 & LEMAS-TTS w/ prosody              & \textbf{15.38} & 0.531 \\
\hline
\multirow{3}{*}{zh} 
 & OpenAudio-S1-Mini                 & 9.59 & 0.550 \\
 & LEMAS-TTS w/o prosody             & 4.64 & \textbf{0.570} \\
 & LEMAS-TTS w/ prosody              & \textbf{3.00} & 0.546 \\
\hline
\multirow{3}{*}{Avg} 
 & OpenAudio-S1-Mini                 & 12.27 & 0.480 \\
 & LEMAS-TTS w/o prosody             & 8.06 & \textbf{0.547} \\
 & LEMAS-TTS w/ prosody              & \textbf{6.39} & 0.539 \\
\hline
\hline
\end{tabular}
\end{table}

As shown in Table~\ref{tab:cmp_wer}, LEMAS-TTS consistently outperforms OpenAudio-S1-mini across all languages, achieving lower WER and higher speaker similarity. We further compare two variants of LEMAS-TTS, with and without a prosody encoder. The prosody-aware model consistently yields lower WER, indicating improved pronunciation stability, while slightly reducing speaker similarity. Subjective listening suggests that prosodic conditioning trades expressive richness for increased articulation stability. We therefore release both variants to accommodate different application preferences. For Vietnamese, the WER of OpenAudio-S1-mini remains unusually high across repeated measurements; we therefore exclude this result from the average WER calculation to avoid disproportionate influence on the aggregated results.

\subsection{Evaluation of LEMAS-Edit}

VoiceCraft~\cite{voicecraft} was originally trained exclusively on the English GigaSpeech corpus. In contrast, LEMAS-Edit is trained on curated subsets from the LEMAS-Dataset, WenetSpeech4TTS, GigaSpeech, and MLS. The resulting combined training corpus covers seven major languages (Zh, En, De, Fr, Pt, Es, It), enabling the model to learn cross-lingual phonotactics and prosody transfer. To efficiently adapt to this multilingual setting, we adopt a warm-start initialization strategy by initializing LEMAS-Edit from the pre-trained 330M-parameter VoiceCraft checkpoint. This design allows LEMAS-Edit to extend VoiceCraft to multilingual speech editing without altering the model architecture.

\begin{figure}[h]
    \centering
    \includegraphics[width=0.9\linewidth]{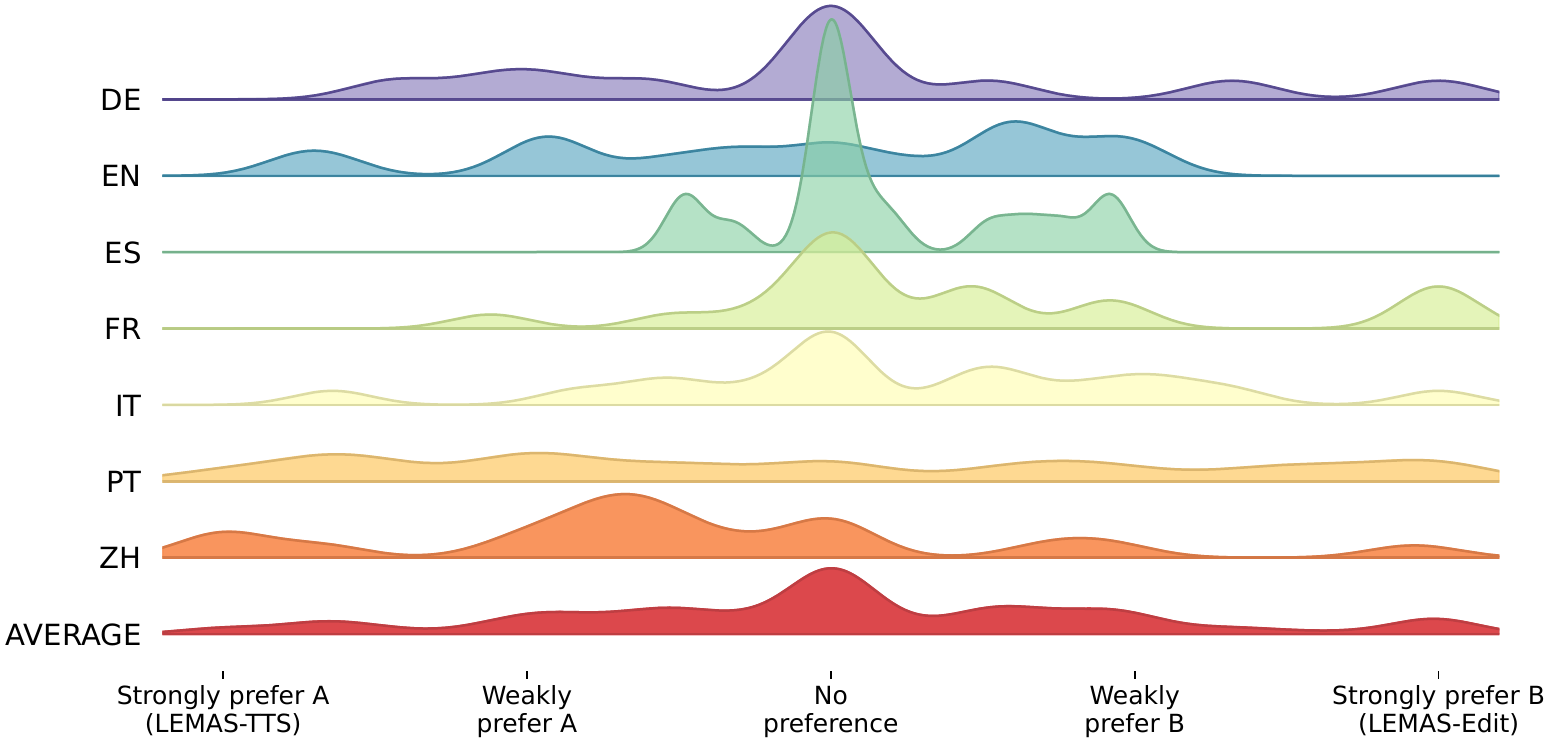}
    \caption{\textbf{Preference distribution of audio naturalness across different languages.} 
    The ridgeline plot illustrates the subjective results of the A/B preference test comparing \textit{LEMAS-TTS} (Model A) and \textit{LEMAS-Edit} (Model B). 
    Individual scores are normalized to a scale of 0--100, where 0 represents a strong preference for Model A and 100 indicates a strong preference for Model B. 
    The vertical distribution (Kernel Density Estimation) for each language reveals the consensus among users, with the ``AVERAGE'' row representing the aggregated performance across all tested languages. }
    \label{fig:edit_abx}
\end{figure}

We further evaluate LEMAS-TTS and LEMAS-Edit on the task of text-based speech editing. Twenty utterances are randomly selected from the LEMAS-Dataset evaluation set, spanning seven languages (2–3 utterances per language). For each utterance, precise word-level alignments are provided, and editing cases are generated by randomly replacing a single word or a phrase using ChatGPT. Six human listeners perform an A/B preference test assessing the naturalness of the edited audio (Figure~\ref{fig:edit_abx}). Although minor language-specific preferences are observed, overall listener judgments are balanced, demonstrating that LEMAS provides a robust, architecture-agnostic benchmark for multilingual speech editing evaluation. All evaluation samples are publicly accessible on the demo page.

\section{Conclusion}
\label{sec:conclusion}

In this work, we introduce \textbf{LEMAS}, a 150K-hour Large-scale Extensible Multilingual Audio Suite with generative speech models, providing a unified and extensible foundation for large-scale multilingual generative speech research. With over 150,000 hours of audio across 10 languages, accompanied by high-quality word-level timestamps and confidence scores, LEMAS-Dataset provides a foundation for training and evaluating next-generation speech generative models. We demonstrated the value of \textbf{LEMAS-Dataset} from two complementary perspectives. Through \textbf{LEMAS-TTS}, we showed that carefully designed alignment and disentanglement objectives are essential for stabilizing large-scale multilingual flow-based models. Through \textbf{LEMAS-Edit}, we validated that precise temporal supervision is a key enabler for robust zero-shot speech editing. Together, these results highlight that progress in generative speech modeling is increasingly driven not only by model architecture, but by the scale, diversity, and granularity of the underlying data. We hope that LEMAS will serve as a shared benchmark and a practical resource for the community, catalyzing further research in multilingual speech synthesis, editing, and controllable audio generation. 

\section*{Acknowledgments}

We would like to sincerely thank the following individuals for their invaluable help and insightful discussions.
The names are listed in alphabetical order by last name: Xie Chen, Chen Cheng, Zhen Guo, Jiaxin Huang, Rui Li, Fei Peng, Daniel van Strien, Tiezhen Wang, Baogang Yao, Fei Yu, Xiaoling Zhang, Changyin Zhou.

% \printbibliography

\bibliographystyle{plainnat}   % 或 unsrtnat / abbrvnat
\bibliography{bibliography}

\begin{thebibliography}{44}
\providecommand{\natexlab}[1]{#1}
\providecommand{\url}[1]{\texttt{#1}}
\expandafter\ifx\csname urlstyle\endcsname\relax
  \providecommand{\doi}[1]{doi: #1}\else
  \providecommand{\doi}{doi: \begingroup \urlstyle{rm}\Url}\fi

\bibitem[Anastassiou et~al.(2024)Anastassiou, Chen, Chen, Chen, Chen, Chen, Cong, Deng, Ding, Gao, Gong, Huang, Huang, Huang, Huo, Jia, Li, Li, Li, Li, Li, Li, Liu, Liu, Liu, Liu, Liu, Liu, Lu, Pan, Wang, Wang, Wang, Wei, Wu, Yao, Yang, Yi, Zhang, Zhang, Zhang, Zhang, Zhang, Zhao, Zhong, and Zhuang]{Anastassiou2024SeedTTSAF}
Philip Anastassiou, Jiawei Chen, Jitong Chen, Yuanzhe Chen, Zhuo Chen, Ziyi Chen, Jian Cong, Lelai Deng, Chuang Ding, Lu~Gao, Mingqing Gong, Peisong Huang, Qingqing Huang, Zhiying Huang, Yuanyuan Huo, Dongya Jia, Chumin Li, Feiya Li, Hui Li, Jiaxin Li, Xiaoyang Li, Xingxing Li, Lin Liu, Shouda Liu, Sichao Liu, Xudong Liu, Yuchen Liu, Zhengxi Liu, Lu~Lu, Junjie Pan, Xin Wang, Yuping Wang, Yuxuan Wang, Zhengnan Wei, Jian Wu, Chao Yao, Yifeng Yang, Yuan-Qiu-Qiang Yi, Junteng Zhang, Qidi Zhang, Shuo Zhang, WenJie Zhang, Yang Zhang, Zilin Zhao, Dejian Zhong, and Xiaobin Zhuang.
\newblock Seed-tts: A family of high-quality versatile speech generation models.
\newblock \emph{arXiv preprint arXiv:2406.02430}, 2024.

\bibitem[Ardila et~al.(2020)Ardila, Branson, Davis, Henretty, Kohler, Meyer, Morais, Saunders, Tyers, and Weber]{ardila2019common}
Rosana Ardila, Megan Branson, Kelly Davis, Michael Henretty, Michael Kohler, Josh Meyer, Reuben Morais, Lindsay Saunders, Francis~M Tyers, and Gregor Weber.
\newblock Common voice: A massively-multilingual speech corpus.
\newblock In \emph{Proceedings of the 12th Language Resources and Evaluation Conference}, 2020.

\bibitem[Baevski et~al.(2020)Baevski, Zhou, Mohamed, and Auli]{baevski2020wav2vec}
Alexei Baevski, Yuhao Zhou, Abdelrahman Mohamed, and Michael Auli.
\newblock wav2vec 2.0: A framework for self-supervised learning of speech representations.
\newblock In \emph{Advances in Neural Information Processing Systems}, pages 12449--12460, 2020.

\bibitem[Borsos et~al.(2023)Borsos, Marinier, Vincent, Kharitonov, Pietquin, Sharifi, Roblek, Teboul, Grangier, Tagliasacchi, and Zeghidour]{audiolm}
Zal{\'{a}}n Borsos, Rapha{\"{e}}l Marinier, Damien Vincent, Eugene Kharitonov, Olivier Pietquin, Matthew Sharifi, Dominik Roblek, Olivier Teboul, David Grangier, Marco Tagliasacchi, and Neil Zeghidour.
\newblock Audiolm: {A} language modeling approach to audio generation.
\newblock \emph{{IEEE} {ACM} Trans. Audio Speech Lang. Process.}, 31:\penalty0 2523--2533, 2023.

\bibitem[Casanova et~al.(2024)Casanova, Davis, G{\"o}lge, G{\"o}knar, Gulea, Hart, Aljafari, Meyer, Morais, Olayemi, and Weber]{Casanova2024XTTSAM}
Edresson Casanova, Kelly Davis, Eren G{\"o}lge, G{\"o}rkem G{\"o}knar, Iulian Gulea, Logan Hart, Aya Aljafari, Joshua Meyer, Reuben Morais, Samuel Olayemi, and Julian Weber.
\newblock Xtts: a massively multilingual zero-shot text-to-speech model.
\newblock \emph{arXiv preprint arXiv:2406.04904}, 2024.

\bibitem[{CETUC}()]{alcaim}
{CETUC}.
\newblock {CETUC} {Alcaim}: {E}uropean {P}ortuguese speech dataset.
\newblock URL \url{http://www02.smt.ufrj.br/~igor.quintanilha/alcaim.tar.gz}.
\newblock Accessed: 2025-12-28.

\bibitem[Chen et~al.(2021)Chen, Chai, Wang, Du, Zhang, Weng, Su, Povey, Trmal, Zhang, et~al.]{chen2021gigaspeech}
Guoguo Chen, Shuzhou Chai, Guanbo Wang, Jiayu Du, Wei-Qiang Zhang, Chao Weng, Dan Su, Daniel Povey, Jan Trmal, Junbo Zhang, et~al.
\newblock Gigaspeech: An evolving, multi-domain asr corpus with 10,000 hours of transcribed audio.
\newblock \emph{arXiv preprint arXiv:2106.06909}, 2021.

\bibitem[Chen et~al.(2025)Chen, Wang, Wu, Zhang, Zhou, Liu, Chen, Liu, Wang, Li, et~al.]{valle}
Sanyuan Chen, Chengyi Wang, Yu~Wu, Ziqiang Zhang, Long Zhou, Shujie Liu, Zhuo Chen, Yanqing Liu, Huaming Wang, Jinyu Li, et~al.
\newblock Neural codec language models are zero-shot text to speech synthesizers.
\newblock \emph{IEEE Transactions on Audio, Speech and Language Processing}, 2025.

\bibitem[Chen et~al.(2024)Chen, Niu, Ma, Deng, Wang, Zhao, Yu, and Chen]{chen2024f5}
Yushen Chen, Zhikang Niu, Ziyang Ma, Keqi Deng, Chunhui Wang, Jian Zhao, Kai Yu, and Xie Chen.
\newblock F5-tts: A fairytaler that fakes fluent and faithful speech with flow matching.
\newblock \emph{arXiv preprint arXiv:2410.06885}, 2024.

\bibitem[Du et~al.(2024{\natexlab{a}})Du, Chen, Zhang, Hu, Lu, Yang, Hu, Zheng, Gu, Ma, et~al.]{cosyvoice}
Zhihao Du, Qian Chen, Shiliang Zhang, Kai Hu, Heng Lu, Yexin Yang, Hangrui Hu, Siqi Zheng, Yue Gu, Ziyang Ma, et~al.
\newblock Cosyvoice: A scalable multilingual zero-shot text-to-speech synthesizer based on supervised semantic tokens.
\newblock \emph{arXiv preprint arXiv:2407.05407}, 2024{\natexlab{a}}.

\bibitem[Du et~al.(2024{\natexlab{b}})Du, Chen, Zhang, Hu, Lu, Yang, Hu, Zheng, Gu, Ma, et~al.]{du2024cosyvoice}
Zhihao Du, Qian Chen, Shiliang Zhang, Kai Hu, Heng Lu, Yexin Yang, Hangrui Hu, Siqi Zheng, Yue Gu, Ziyang Ma, et~al.
\newblock Cosyvoice: A scalable multilingual zero-shot text-to-speech synthesizer based on supervised semantic tokens.
\newblock \emph{arXiv preprint arXiv:2407.05407}, 2024{\natexlab{b}}.

\bibitem[Du et~al.(2024{\natexlab{c}})Du, Wang, Chen, Shi, Lv, Zhao, Gao, Yang, Gao, Wang, et~al.]{du2024cosyvoice2}
Zhihao Du, Yuxuan Wang, Qian Chen, Xian Shi, Xiang Lv, Tianyu Zhao, Zhifu Gao, Yexin Yang, Changfeng Gao, Hui Wang, et~al.
\newblock Cosyvoice 2: Scalable streaming speech synthesis with large language models.
\newblock \emph{arXiv preprint arXiv:2412.10117}, 2024{\natexlab{c}}.

\bibitem[Du et~al.(2025)Du, Gao, Wang, Yu, Zhao, Wang, Lv, Wang, Ni, Shi, An, Yang, Li, Chen, Gao, Chen, Gu, Chen, Chen, Zhang, Wang, and Ye]{Du2025CosyVoice3T}
Zhihao Du, Changfeng Gao, Yuxuan Wang, Fan Yu, Tianyu Zhao, Hao Wang, Xiang Lv, Hui Wang, Chongjia Ni, Xian Shi, Keyu An, Guanrou Yang, Yabin Li, Yanni Chen, Zhifu Gao, Qian Chen, Yue Gu, Mengzhe Chen, Yafeng Chen, Shi-Min Zhang, Wen Wang, and Jieping Ye.
\newblock Cosyvoice 3: Towards in-the-wild speech generation via scaling-up and post-training.
\newblock \emph{arXiv preprint arXiv:2505.17589}, 2025.

\bibitem[{eSpeak NG}(2022)]{espeakng}
{eSpeak NG}.
\newblock espeak ng (version 1.51).
\newblock \url{https://github.com/espeak-ng/espeak-ng/tree/1.51}, 2022.
\newblock Accessed: 2025-12-25.

\bibitem[Gales et~al.(2014)Gales, Rath, et~al.]{gales2014speech}
Mark~JF Gales, Shakti~P Rath, et~al.
\newblock Speech recognition and keyword spotting for low-resource languages: Babel project research at cued.
\newblock \emph{Spoken Language Technologies for Under-Resourced Languages}, 2014.

\bibitem[Ganin and Lempitsky(2015)]{grl}
Yaroslav Ganin and Victor Lempitsky.
\newblock Unsupervised domain adaptation by backpropagation.
\newblock In \emph{International conference on machine learning}, pages 1180--1189. PMLR, 2015.

\bibitem[Graves et~al.(2006)Graves, Fern{\'a}ndez, Gomez, and Schmidhuber]{graves2006ctc}
Alex Graves, Santiago Fern{\'a}ndez, Faustino Gomez, and J{\"u}rgen Schmidhuber.
\newblock Connectionist temporal classification: labelling unsegmented sequence data with recurrent neural networks.
\newblock In \emph{Proceedings of the 23rd international conference on Machine learning}, pages 369--376, 2006.

\bibitem[Guo et~al.(2024)Guo, Hu, Liu, Shen, Tang, Wu, Xie, Xie, and Xu]{guo2024fireredtts}
Hao-Han Guo, Yao Hu, Kun Liu, Fei-Yu Shen, Xu~Tang, Yi-Chen Wu, Feng-Long Xie, Kun Xie, and Kai-Tuo Xu.
\newblock Fireredtts: A foundation text-to-speech framework for industry-level generative speech applications.
\newblock \emph{arXiv preprint arXiv:2409.03283}, 2024.

\bibitem[He et~al.(2024)He, Shang, Wang, Li, Gu, Hua, Liu, Yang, Li, Shi, et~al.]{he2024emilia}
Haorui He, Zengqiang Shang, Chaoren Wang, Xuyuan Li, Yicheng Gu, Hua Hua, Liwei Liu, Chen Yang, Jiaqi Li, Peiyang Shi, et~al.
\newblock Emilia: An extensive, multilingual, and diverse speech dataset for large-scale speech generation.
\newblock In \emph{2024 IEEE Spoken Language Technology Workshop (SLT)}, pages 885--890. IEEE, 2024.

\bibitem[Kahn et~al.(2020)Kahn, Riviere, Zheng, Kharitonov, Xu, Mazar{\'e}, Karadayi, Likhomanenko, Pratap, Kahn, et~al.]{kahn2020libri}
Jacob Kahn, Morgane Riviere, Weiyi Zheng, Eugene Kharitonov, Qiantong Xu, Pierre-Emmanuel Mazar{\'e}, Julien Karadayi, Vitaliy Likhomanenko, Vineel Pratap, Awni Kahn, et~al.
\newblock Libri-light: A benchmark for asr with limited or no supervision.
\newblock In \emph{IEEE International Conference on Acoustics, Speech and Signal Processing (ICASSP)}, pages 7669--7673, 2020.

\bibitem[Karpov et~al.(2021)Karpov, Denisenko, and Minkin]{golos}
Nikolay Karpov, Alexander Denisenko, and Fedor Minkin.
\newblock Golos: Russian dataset for speech research.
\newblock \emph{arXiv preprint arXiv:2106.10161}, 2021.

\bibitem[Kim et~al.(2021)Kim, Kong, and Son]{vits}
Jaehyeon Kim, Jungil Kong, and Juhee Son.
\newblock Conditional variational autoencoder with adversarial learning for end-to-end text-to-speech.
\newblock In \emph{International Conference on Machine Learning}, pages 5530--5540. PMLR, 2021.

\bibitem[Le et~al.(2023)Le, Vyas, Shi, Karrer, Sari, Moritz, Williamson, Manohar, Adi, Mahadeokar, et~al.]{le2023voicebox}
Matthew Le, Apoorv Vyas, Bowen Shi, Brian Karrer, Leda Sari, Rashel Moritz, Mary Williamson, Vimal Manohar, Yossi Adi, Jay Mahadeokar, et~al.
\newblock Voicebox: Text-guided multilingual universal speech generation at scale.
\newblock \emph{Advances in neural information processing systems}, 36:\penalty0 14005--14034, 2023.

\bibitem[Li et~al.(2023)Li, Takamichi, Saeki, Chen, Shiota, and Watanabe]{li2023yodas}
Xinjian Li, Shinnosuke Takamichi, Takaaki Saeki, William Chen, Sayaka Shiota, and Shinji Watanabe.
\newblock Yodas: Youtube-oriented dataset for audio and speech.
\newblock In \emph{2023 IEEE Automatic Speech Recognition and Understanding Workshop (ASRU)}, pages 1--8. IEEE, 2023.

\bibitem[Ma et~al.(2024)Ma, Guo, Song, Jiang, Wang, Xue, Xu, Zhao, Zhang, and Xie]{ma2024wenetspeech4tts}
Linhan Ma, Dake Guo, Kun Song, Yuepeng Jiang, Shuai Wang, Liumeng Xue, Weiming Xu, Huan Zhao, Binbin Zhang, and Lei Xie.
\newblock Wenetspeech4tts: A 12,800-hour mandarin tts corpus for large speech generation model benchmark.
\newblock \emph{arXiv preprint arXiv:2406.05763}, 2024.

\bibitem[McAuliffe et~al.(2017)McAuliffe, Socolof, Mihuc, Wagner, and Sonderegger]{montreal-forced-aligner}
M.~McAuliffe, M.~Socolof, S.~Mihuc, M.~Wagner, and M.~Sonderegger.
\newblock Montreal forced aligner: Trainable text-speech alignment using kaldi.
\newblock In \emph{Proc. Interspeech 2017}, pages 498--502, 2017.

\bibitem[Panayotov et~al.(2015)Panayotov, Chen, Povey, and Khudanpur]{panayotov2015librispeech}
Vassil Panayotov, Guoguo Chen, Daniel Povey, and Sanjeev Khudanpur.
\newblock Librispeech: An asr corpus based on public domain audio books.
\newblock In \emph{IEEE International Conference on Acoustics, Speech and Signal Processing (ICASSP)}, 2015.

\bibitem[Peng et~al.(2024)Peng, Huang, Li, Mohamed, and Harwath]{voicecraft}
Puyuan Peng, Po-Yao Huang, Shang-Wen Li, Abdelrahman Mohamed, and David Harwath.
\newblock Voicecraft: Zero-shot speech editing and text-to-speech in the wild.
\newblock \emph{arXiv preprint arXiv:2403.16973}, 2024.

\bibitem[Pratap et~al.(2020)Pratap, Xu, Sriram, Synnaeve, and Collobert]{pratap2020mls}
Vineel Pratap, Qiantong Xu, Anuroop Sriram, Gabriel Synnaeve, and Ronan Collobert.
\newblock Mls: A large-scale multilingual dataset for speech research.
\newblock In \emph{Interspeech}, 2020.

\bibitem[Pratap et~al.(2023)Pratap, Tjandra, Shi, Tomasello, Babu, Kundu, Elkahky, Ni, Vyas, et~al.]{pratap2023scaling}
Vineel Pratap, Andros Tjandra, Bowen Shi, Paden Tomasello, Arun Babu, Sravya Kundu, Ali Elkahky, Zhaoheng Ni, Apoorv Vyas, et~al.
\newblock Scaling speech technology to 1,000+ languages.
\newblock \emph{Journal of Machine Learning Research}, 24\penalty0 (368), 2023.

\bibitem[Ren et~al.(2019)Ren, Ruan, Tan, Qin, Zhao, Zhao, and Liu]{fastspeech}
Yi~Ren, Yangjun Ruan, Xu~Tan, Tao Qin, Sheng Zhao, Zhou Zhao, and Tie-Yan Liu.
\newblock Fastspeech: Fast, robust and controllable text to speech.
\newblock volume~32, 2019.

\bibitem[Salesky et~al.(2021)Salesky, Wiesner, Bremerman, Cattoni, Negri, Turchi, Oard, and Post]{tedx}
Elizabeth Salesky, Matthew Wiesner, Jacob Bremerman, Roldano Cattoni, Matteo Negri, Marco Turchi, Douglas~W. Oard, and Matt Post.
\newblock The multilingual tedx corpus for speech recognition and translation.
\newblock \emph{arXiv preprint arXiv:2102.01757}, 2021.

\bibitem[Schultz and Schlippe(2013)]{schultz2013globalphone}
Tanja Schultz and Tim Schlippe.
\newblock Globalphone: A multilingual text \& speech database in 20 languages.
\newblock In \emph{IEEE International Conference on Acoustics, Speech and Signal Processing}, 2013.

\bibitem[{Seamless Communication} et~al.(2023){Seamless Communication}, Barrault, Chung, Meglioli, Dale, Dong, Duquenne, Elsahar, Gong, Heffernan, Hoffman, Klaiber, Li, Licht, Maillard, Rakotoarison, Sadagopan, Wenzek, Ye, Akula, Chen, Hachem, Ellis, Gonzalez, Haaheim, Hansanti, Howes, Huang, Hwang, Inaguma, Jain, Kalbassi, Kallet, Kulikov, Lam, Li, Ma, Mavlyutov, Peloquin, Ramadan, Ramakrishnan, Sun, Tran, Tran, Tufanov, Vogeti, Wood, Yang, Yu, Andrews, Balioglu, Costa-jussà, Celebi, Elbayad, Gao, Guzmán, Kao, Lee, Mourachko, Pino, Popuri, Ropers, Saleem, Schwenk, Tomasello, Wang, Wang, and Wang]{SeamlessM4TArXiv}
{Seamless Communication}, Loïc Barrault, Yu-An Chung, Mariano~Cora Meglioli, David Dale, Ning Dong, Paul-Ambroise Duquenne, Hady Elsahar, Hongyu Gong, Kevin Heffernan, John Hoffman, Christopher Klaiber, Pengwei Li, Daniel Licht, Jean Maillard, Alice Rakotoarison, Kaushik~Ram Sadagopan, Guillaume Wenzek, Ethan Ye, Bapi Akula, Peng-Jen Chen, Naji~El Hachem, Brian Ellis, Gabriel~Mejia Gonzalez, Justin Haaheim, Prangthip Hansanti, Russ Howes, Bernie Huang, Min-Jae Hwang, Hirofumi Inaguma, Somya Jain, Elahe Kalbassi, Amanda Kallet, Ilia Kulikov, Janice Lam, Daniel Li, Xutai Ma, Ruslan Mavlyutov, Benjamin Peloquin, Mohamed Ramadan, Abinesh Ramakrishnan, Anna Sun, Kevin Tran, Tuan Tran, Igor Tufanov, Vish Vogeti, Carleigh Wood, Yilin Yang, Bokai Yu, Pierre Andrews, Can Balioglu, Marta~R. Costa-jussà, Onur Celebi, Maha Elbayad, Cynthia Gao, Francisco Guzmán, Justine Kao, Ann Lee, Alexandre Mourachko, Juan Pino, Sravya Popuri, Christophe Ropers, Safiyyah Saleem, Holger Schwenk, Paden Tomasello, Changhan Wang, Jeff
  Wang, and Skyler Wang.
\newblock Seamlessm4t-massively multilingual \& multimodal machine translation, 2023.

\bibitem[Shen et~al.(2024)Shen, Ju, Tan, Liu, Leng, He, Qin, Zhao, and Bian]{shen2023naturalspeech2}
Kai Shen, Zeqian Ju, Xu~Tan, Eric Liu, Yichong Leng, Lei He, Tao Qin, Sheng Zhao, and Jiang Bian.
\newblock Naturalspeech 2: Latent diffusion models are natural and zero-shot speech and singing synthesizers.
\newblock In \emph{The Twelfth International Conference on Learning Representations}, 2024.

\bibitem[Team(2024)]{fishaudio}
Fish~Audio Team.
\newblock Fish speech: Openaudio s1 mini.
\newblock \url{https://huggingface.co/fishaudio/openaudio-s1-mini}, 2024.

\bibitem[Wang et~al.(2021)Wang, Riviere, Lee, Wu, Talwalkar, Xiao, Saraf, Puvvada, and Dupoux]{wang2021voxpopuli}
Changhan Wang, Morgane Riviere, Ann Lee, Anne Wu, Chaitanya Talwalkar, Elizabeth Xiao, Yatharth Saraf, Piyush Puvvada, and Emmanuel Dupoux.
\newblock Voxpopuli: A large-scale multilingual speech corpus for representation learning, semi-supervised learning and interpretation.
\newblock In \emph{Association for Computational Linguistics (ACL)}, 2021.

\bibitem[Wang et~al.(2025)Wang, Jiang, Ma, Zhang, Liu, Li, Liang, Zheng, Wang, Feng, et~al.]{wang2025spark}
Xinsheng Wang, Mingqi Jiang, Ziyang Ma, Ziyu Zhang, Songxiang Liu, Linqin Li, Zheng Liang, Qixi Zheng, Rui Wang, Xiaoqin Feng, et~al.
\newblock Spark-tts: An efficient llm-based text-to-speech model with single-stream decoupled speech tokens.
\newblock \emph{arXiv preprint arXiv:2503.01710}, 2025.

\bibitem[Wang et~al.(2024)Wang, Zhan, Liu, Zeng, Guo, Zheng, Zhang, Zhang, Zhang, and Wu]{Wang2024MaskGCTZT}
Yuancheng Wang, Haoyue Zhan, Liwei Liu, Ruihong Zeng, Haotian Guo, Jiachen Zheng, Qiang Zhang, Xueyao Zhang, Shunsi Zhang, and Zhizheng Wu.
\newblock Maskgct: Zero-shot text-to-speech with masked generative codec transformer.
\newblock \emph{arXiv preprint arXiv:2409.00750}, 2024.

\bibitem[Wang et~al.(2017)Wang, Skerry-Ryan, Stanton, Wu, Weiss, Jaitly, Yang, Xiao, Chen, Bengio, et~al.]{tacotron}
Yuxuan Wang, RJ~Skerry-Ryan, Daisy Stanton, Yonghui Wu, Ron~J Weiss, Navdeep Jaitly, Zongheng Yang, Ying Xiao, Zhifeng Chen, Samy Bengio, et~al.
\newblock Tacotron: Towards end-to-end speech synthesis.
\newblock 2017.

\bibitem[Xie et~al.(2025)Xie, Shen, Li, Xie, Tang, and Hu]{Xie2025FireRedTTS2TL}
Kun Xie, Feiyu Shen, Junjie Li, Fenglong Xie, Xu~Tang, and Yao Hu.
\newblock Fireredtts-2: Towards long conversational speech generation for podcast and chatbot.
\newblock \emph{arXiv preprint arXiv:2509.02020}, 2025.

\bibitem[Yang et~al.(2025)Yang, Song, Zhuo, Cui, Li, Yang, Du, Ma, Liu, Wang, et~al.]{yang2025gigaspeech}
Yifan Yang, Zheshu Song, Jianheng Zhuo, Mingyu Cui, Jinpeng Li, Bo~Yang, Yexing Du, Ziyang Ma, Xunying Liu, Ziyuan Wang, et~al.
\newblock Gigaspeech 2: An evolving, large-scale and multi-domain asr corpus for low-resource languages with automated crawling, transcription and refinement.
\newblock In \emph{Proceedings of the 63rd Annual Meeting of the Association for Computational Linguistics (Volume 1: Long Papers)}, pages 2673--2686, 2025.

\bibitem[Zhang et~al.(2022)Zhang, Lv, Guo, Shao, Yang, Xie, Xu, Bu, Chen, Zeng, et~al.]{zhang2022wenetspeech}
Binbin Zhang, Hang Lv, Pengcheng Guo, Qijie Shao, Chao Yang, Lei Xie, Xin Xu, Hui Bu, Xiaoyu Chen, Chenchen Zeng, et~al.
\newblock Wenetspeech: A 10000+ hours multi-domain mandarin corpus for speech recognition.
\newblock In \emph{ICASSP 2022-2022 IEEE International Conference on Acoustics, Speech and Signal Processing (ICASSP)}, pages 6182--6186. IEEE, 2022.

\bibitem[Zhou et~al.(2025)Zhou, Zhou, He, Zhou, Wang, Deng, and Shu]{indextts2}
Siyi Zhou, Yiquan Zhou, Yi~He, Xun Zhou, Jinchao Wang, Wei Deng, and Jingchen Shu.
\newblock Indextts2: A breakthrough in emotionally expressive and duration-controlled auto-regressive zero-shot text-to-speech.
\newblock \emph{arXiv preprint arXiv:2506.21619}, 2025.

\end{thebibliography}

\newpage

\appendix
\addcontentsline{toc}{section}{Appendices}
\addtocontents{toc}{\protect\setcounter{tocdepth}{0}}
\newgeometry{left=0.5cm, right=0.5cm, bottom=0.5cm, top=0.5cm}

\restoregeometry
\end{document}